\documentclass[amsmath,amssymb,showpacs]{revtex4}
\usepackage{epsfig}
\usepackage{graphicx}

\begin{document}

\title{Gravitational wave in Lorentz violating gravity}

\author{Xin Li}
\email{lixin@ihep.ac.cn}
\author{Zhe Chang}
\email{changz@ihep.ac.cn}
\affiliation{Institute of High Energy Physics\\
and\\
Theoretical Physics Center for Science Facilities,\\
Chinese Academy of Sciences, 100049 Beijing, China}

\begin{abstract}
By making use of the weak gravitational field approximation, we obtain a linearized solution of the gravitational vacuum field equation in an anisotropic spacetime. The plane-wave solution and dispersion relation of gravitational wave is presented explicitly.
There is possibility that the speed of gravitational wave is larger than the speed of light and the casuality still holds.
We show that the energy-momentum of gravitational wave in the ansiotropic spacetime is still well defined and conserved.

\end{abstract}

\maketitle
\section{Introduction}
Lorentz Invariance is one of the foundations of the Standard
model of particle physics. The constraints on possible Lorentz violating phenomenology are quite severe, see for example, the summary tables that provided by Kostelecky {\it et al}.\cite{Kostelecky1}. The gravitational interaction is far more weak, compare to other fundamental interactions. This allows one to study the possible Lorentz violating effects on certain gravity theories, such as Einstein-aether theory \cite{Jacobson} and Horava-Lifshitz theory \cite{Horava}. One feature of Lorentz invariance violation is that the speed of light differ from the one in special relativity. The gravity theories with Lorentz violation could have the feature that the speed of graviton or the speed of gravitational wave differ from the one in general relativity. Studying the speed of gravitational wave in a Lorentz violating gravity theory will give different perspective on quantum gravitational phenomena.

One of the most important prediction of Einstein's general relativity is gravitational radiation. Many pioneer works \cite{Braginsky,Thorne,Weiss} have discussed the gravitational radiation in both theoretical properties and experimental approaches of detections. Currently, the most sensitive measurement is provided by ground-based Laser Interferometer Gravitational-Wave Observatory (LIGO) detector \cite{LIGO}. Another sensitive measurement, which is in progress, is the Laser Interferometer Space Antenna (LISA) that detect and accurately measure gravitational waves from astronomical sources.
The primordial gravitational waves \cite{Krauss,Grishchuk} could be of interest to cosmologists as they provide a new and unique window on the earliest moments in the history of the universe and on possible new physics at energies many orders of magnitude beyond those accessible at particle accelerators.

In general relativity, the effects of gravitation are ascribed to spacetime curvature instead of a force.
However, up to now, general relativity still faces problems.

First, the recent astronomical observations \cite{Riess} found that our universe is accelerated expanding. This result can not be obtained directly from Einstein's gravity and his cosmological principle. Since normal matters only provide attractive force. The most widely adopted way to resolve it is involving the so called dark energy which provides the repulsive force.

Second,  the flat rotation curves of spiral galaxies violate the
prediction of Einstein's gravity \cite{Zwicky}. The most widely adopted way to resolve it is involving the so called dark matter which provides enough attractive force such that the discrepancy is restored.

The above astronomical phenomena occur at very large cosmological scale. The following anomalies occur in solar system which imply the Newton's inverse-square law of universal gravitation and general relativity need modifications.

The third one, two Pioneer spacecrafts suffer an anomalous constant sunward acceleration, $a_p=(8.74\pm1.33)\times 10^{-10}{\rm m/s^2}$ \cite{Anderson}.

The fourth one, it has been observed at various occasions that satellites after an Earth swing-by possess a significant
unexplained velocity increase by a few mm/s \cite{Anderson1}.

The fifth one, from the analysis of radiometric measurements of distances between the Earth and the major planets
including observations from Martian orbiters and landers from 1961 to 2003 a secular trend of
the Astronomical Unit of $15\pm4$ m/cy has been reported \cite{Krasinsky}.

The sixth one, a recent orbital analysis of Lunar Laser Ranging (LLR) \cite{Williams} shows an anomalous secular eccentricity variation of the Moon's orbit $\rm (0.9\pm0.3)\times10^{-11}/yr$.

All the facts imply that the Einstein's theory should be modified.
By mimicking Einstein, we have proposed that the modified gravitational theory
should correspond to a new geometry which involves Riemann
geometry as its special case.  Finsler geometry \cite{Book by Bao} as a nature extension of Riemann geometry
is a good candidate to solve the problems mentioned above.
A new geometry (Finsler geometry) involves new spacetime symmetry. The Lorentz violation is intimately linked to Finsler geometry. Kostelecky \cite{Kostelecky} have studied effective field theories with explicit Lorentz violation in Finsler spacetime.

Finsler geometry really gives better description for the
nature of gravity: the
flat rotation curves of spiral galaxies can be deduced naturally
without invoking dark matter \cite{Finsler DM}; a Finlerian gravity model could account for the accelerated expanding university without invoking dark energy \cite{Finsler DE}; a special Finsler space-Randers
space \cite{Finsler PA} could account for the anomalous
acceleration \cite{Anderson} in solar system observed by Pioneer 10
and 11 spacecrafts; the Finsler spacetime could give a modification on the gravitational deflection of light \cite {Finsler BL}, which may account to these observations without adding dark matter in Bullet Cluster \cite{Clowe}; the result based on the kinematics with a special Finsler spacetime is in good agreement with secular trend of the Astronomical Unit and secular eccentricity variation of the Moon's orbit \cite{Finsler AU}.

It is interest to investigate the gravitational wave in Finsler spacetime. It is well known that the gravitational wave propagates with the speed of light in general relativity. This is due to the fact that the spacetime metric is close to the Minkowski metric in the weak gravitational field approximation, and the causal speed of Minkowski spacetime is just the speed of light. However, in Finsler spacetime the causal speed is generally different with the speed of light \cite{Pfeifer}.

In this paper, we will present the solution of linearized gravitational vacuum field equation in Finsler spacetime. It is shown that there is possibility that the causal speed of it is larger than the speed of light.

\section{Vacuum field equation in Finsler spacetime}
Instead of defining an inner product structure over the tangent bundle in Riemann geometry, Finsler geometry is based on
the so called Finsler structure $F$ with the property
$F(x,\lambda y)=\lambda F(x,y)$ for all $\lambda>0$, where $x$ represents position
and $y\equiv\frac{dx}{d\tau}$ represents velocity. The Finsler metric is given as \cite{Book
by Bao}
 \begin{equation}
 g_{\mu\nu}\equiv\frac{\partial}{\partial
y^\mu}\frac{\partial}{\partial y^\nu}\left(\frac{1}{2}F^2\right).
\end{equation}
Finsler geometry has its genesis in integrals of the form
\begin{equation}
\label{integral length}
\int^r_sF(x^1,\cdots,x^n;\frac{dx^1}{d\tau},\cdots,\frac{dx^n}{d\tau})d\tau~.
\end{equation}
The Finsler structure represents the length element of Finsler space.

The parallel transport
has been studied in the framework of Cartan
connection \cite{Matsumoto,Antonelli,Szabo}. The notation of parallel
transport in Finsler manifold means that the length
$F\left(\frac{dx}{d\tau}\right)$ is constant.
The geodesic equation for Finsler manifold is given as \cite{Book by Bao}
\begin{equation}
\label{geodesic}
\frac{d^2x^\mu}{d\tau^2}+2G^\mu=0,
\end{equation}
where
\begin{equation}
\label{geodesic spray}
G^\mu=\frac{1}{4}g^{\mu\nu}\left(\frac{\partial^2 F^2}{\partial x^\lambda \partial y^\nu}y^\lambda-\frac{\partial F^2}{\partial x^\nu}\right)
\end{equation} is called geodesic spray coefficient.
Obviously, if $F$ is Riemannian metric, then
\begin{equation}
G^\mu=\frac{1}{2}\tilde{\gamma}^\mu_{\nu\lambda}y^\nu y^\lambda,
\end{equation}
where $\tilde{\gamma}^\mu_{\nu\lambda}$ is the Riemannian Christoffel symbol.
Since the geodesic equation (\ref{geodesic}) is directly
derived from the integral length
\begin{equation} L=\int
F\left(\frac{dx}{d\tau}\right)d\tau,
\end{equation} the inner product
$\left(\sqrt{g_{\mu\nu}\frac{dx^\mu}{d\tau}\frac{dx^\nu}{d\tau}}=F\left(\frac{dx}{d\tau}\right)\right)$
of two parallel transported vectors is preserved.

In Finsler manifold, there exists a linear connection~-~the
Chern connection \cite{Chern}. It is torsion freeness and almost
metric-compatibility,
 \begin{equation}\label{Chern connection}
 \Gamma^{\alpha}_{\mu\nu}=\gamma^{\alpha}_{\mu\nu}-g^{\alpha\lambda}\left(A_{\lambda\mu\beta}\frac{N^\beta_\nu}{F}-A_{\mu\nu\beta}\frac{N^\beta_\lambda}{F}+A_{\nu\lambda\beta}\frac{N^\beta_\mu}{F}\right),
 \end{equation}
 where $\gamma^{\alpha}_{\mu\nu}$ is the formal Christoffel symbols of the
second kind with the same form of Riemannian connection, $N^\mu_\nu$
is defined as
$N^\mu_\nu\equiv\gamma^\mu_{\nu\alpha}y^\alpha-A^\mu_{\nu\lambda}\gamma^\lambda_{\alpha\beta}y^\alpha
y^\beta$
 and $A_{\lambda\mu\nu}\equiv\frac{F}{4}\frac{\partial}{\partial y^\lambda}\frac{\partial}{\partial y^\mu}\frac{\partial}{\partial y^\nu}(F^2)$ is the
 Cartan tensor (regarded as a measurement of deviation from the Riemannian
 Manifold). In terms of Chern connection, the curvature of Finsler space is given as
\begin{equation}\label{Finsler curvature}
R^{~\lambda}_{\kappa~\mu\nu}=\frac{\delta
\Gamma^\lambda_{\kappa\nu}}{\delta x^\mu}-\frac{\delta
\Gamma^\lambda_{\kappa\mu}}{\delta
x^\nu}+\Gamma^\lambda_{\alpha\mu}\Gamma^\alpha_{\kappa\nu}-\Gamma^\lambda_{\alpha\nu}\Gamma^\alpha_{\kappa\mu},
\end{equation}
where $\frac{\delta}{\delta x^\mu}=\frac{\partial}{\partial x^\mu}-N^\nu_\mu\frac{\partial}{\partial y^\nu}$.

The gravity in Finsler spacetime has been investigated for a long
time \cite{Takano,Ikeda,Tavakol1, Bogoslovsky1}. In this paper, we introduce vacuum field equation by the way discussed first by Pirani \cite{Pirani, Rutz}. In Newton's theory of gravity, the equation of motion of a test particle is given as
\begin{equation}
\label{dynamic Newton}
\frac{d^2x^i}{dt^2}=-\eta^{ij}\frac{\partial \phi}{\partial x^i},
\end{equation}
where $\phi=\phi(x)$ is the gravitational potential and $\eta^{ij}$ is Euclidean metric. For an infinitesimal transformation $x^i\rightarrow x^i+\epsilon\xi^i$($|\epsilon|\ll1$), the equation (\ref{dynamic Newton}) becomes, up to first order in $\epsilon$,
\begin{equation}
\label{dynamic Newton1}
\frac{d^2x^i}{dt^2}+\epsilon\frac{d^2\xi^i}{dt^2}=-\eta^{ij}\frac{\partial \phi}{\partial x^i}-\epsilon\eta^{ij}\xi^k\frac{\partial^2\phi}{\partial x^j\partial x^k}.
\end{equation}
Combining the above equations(\ref{dynamic Newton}) and (\ref{dynamic Newton1}), we obtain
\begin{equation}
\frac{d^2\xi^i}{dt^2}=\eta^{ij}\xi^k\frac{\partial^2\phi}{\partial x^j\partial x^k}\equiv\xi^kH^i_k.
\end{equation}
In Newton's theory of gravity, the vacuum field equation is given as $H^i_i=\bigtriangledown^2\phi=0$. It means that the tensor $H^i_k$ is traceless in Newton's vacuum.

In general relativity, the geodesic deviation gives similar equation
\begin{equation}
\frac{D^2\xi^\mu}{D\tau^2}=\xi^\nu \tilde{R}^\mu_{~\nu},
\end{equation}
where $\tilde{R}^\mu_{~\nu}=\tilde{R}^{~\mu}_{\lambda~\nu\rho}\frac{dx^\lambda}{d\tau}\frac{dx^\rho}{d\tau}$. Here, $\tilde{R}^{~\mu}_{\lambda~\nu\rho}$ is Riemannian curvature tensor, $D$ denotes the covariant derivative alone the curve $x^\mu(\tau)$. The vacuum field equation in general relativity gives $\tilde{R}^{~\lambda}_{\mu~\lambda\nu}=0$\cite{Weinberg}. It implies that the tensor $\tilde{R}^\mu_{~\nu}$ is also traceless, $\tilde{R}\equiv\tilde{R}^\mu_{~\mu}=0$.

In Finsler spacetime, the geodesic deviation gives \cite{Book by Bao}
\begin{equation}
\frac{D^2\xi^\mu}{D\tau^2}=\xi^\nu R^\mu_{~\nu},
\end{equation}
where $R^\mu_{~\nu}=R^{~\mu}_{\lambda~\nu\rho}\frac{dx^\lambda}{d\tau}\frac{dx^\rho}{d\tau}$. Here, $R^{~\mu}_{\lambda~\nu\rho}$ is Finsler curvature tensor defined in (\ref{Finsler curvature}), $D$ denotes covariant derivative $\frac{D\xi^\mu}{D\tau}=\frac{d\xi^\mu}{d\tau}+\xi^\nu\frac{dx^\lambda}{d\tau}\Gamma^\mu_{\nu\lambda}(x,\frac{dx}{d\tau})$. Since the vacuum field equations of Newton's gravity and general relativity have similar form, we may assume that vacuum field equation in Finsler spacetime hold similar requirement as the case of Netwon's gravity and general relativity. It implies that the tensor $R^\mu_{~\nu}$ in Finsler geodesic deviation equation should be traceless, $R\equiv R^\mu_{~\mu}=0$. We have proved that the analogy from the geodesic deviation equation is valid at least in Finsler spacetime of Berwald type \cite{Finsler DM}. For this reason, we may suppose that this analogy is valid in general Finsler spacetime.

It should be noticed that $H$ is called the Ricci scaler, which is a geometrical invariant. For a tangent plane $\Pi\subset T_xM$ and a non-zero vector $y\in T_xM$, the flag curvature is defined as
\begin{equation}
\label{flag curvature}
K(\Pi,y)\equiv\frac{g_{\lambda\mu}R^\mu_{~\nu}u^\nu u^\lambda}{F^2g_{\rho\theta}u^\rho u^\theta-(g_{\sigma\kappa}y^\sigma u^\kappa)^2},
\end{equation}
where $u\in\Pi$. The flag curvature is a geometrical invariant that generalizes the sectional curvature of Riemannian geometry. It is clear that the Ricci scaler $R$ is the trace of $R^\mu_{~\nu}$, which is the predecessor of flag curvature. Therefore, the value of Ricci scaler $R$ is invariant under the ordinate transformation. Furthermore, the predecessor of flag curvature could be written in terms of the geodesic spray coefficient
\begin{equation}\label{predecessor flag curvature}
R^\mu_{~\nu}=2\frac{\partial G^\mu}{\partial x^\nu}-y^\lambda\frac{\partial^2 G^\mu}{\partial x^\lambda\partial y^\nu}+2G^\lambda\frac{\partial^2 G^\mu}{\partial y^\lambda\partial y^\nu}-\frac{\partial G^\mu}{\partial y^\lambda}\frac{\partial G^\lambda}{\partial y^\nu}.
\end{equation}
Thus, the Ricci scaler $R$ is insensitive to connection that one is using, it only depends on the length element $F$. The gravitational vacuum field equation $R=0$ is universal in any types of theories of Finsler gravity. Pfeifer {\it et al}. \cite{Pfeifer1} have constructed gravitational dynamics for Finsler spacetimes in terms of an action integral on the unit tangent bundle. Their researches also show that the gravitational vacuum field equation in Finsler spacetime is $R=0$.

\section{Gravitational wave in Finslerian vacuum}
It is hard to find a non trivial solution of the gravitational vacuum field equation ($R=0$) in Finsler spacetime. Here, we study the weak field radiative solution of the Finslerian vacuum field equation $R=0$. It is well known that the Minkowski spacetime is a trivial solution of Einstein's vacuum field equation. In the Finsler spacetime, the trivial solution of Finslerian vacuum field equation is called locally Minkowski spacetime. A Finsler spacetime is called a locally Minkowshi spacetime if there is a local coordinate system $(x^\mu)$, with induced tangent space coordinates $y^\mu$, such that $F$ depends only on $y$ and not on $x$. Using the formula (\ref{predecessor flag curvature}), one knows obvious that locally Minkowski spacetime is a solution of Finslerian vacuum field equation.

We suppose that the metric is close to the locally Minkowski metric $\eta_{\mu\nu}(y)$,
\begin{equation}
g_{\mu\nu}=\eta_{\mu\nu}(y)+h_{\mu\nu}(x,y),
\end{equation}
where $|h_{\mu\nu}|\ll1$. In the rest of this section, the lowering and raising of indices are carried out by $\eta_{\mu\nu}$ and its matrix inverse $\eta^{\mu\nu}$.
To first order in $h$, the geodesic spray coefficient is
\begin{equation}
G^\mu=\frac{1}{4}\eta^{\mu\nu}\left(2\frac{\partial h_{\alpha\nu}}{\partial x^\lambda}y^\alpha y^\lambda-\frac{\partial h_{\alpha\beta}}{\partial x^\nu}y^\alpha y^\beta\right).
\end{equation}
We have already used the Euler's theorem for homogeneous functions to obtain the above equation.
And the Ricci scaler is
\begin{eqnarray}
R=R^\mu_{~\mu}&=&2\frac{\partial G^\mu}{\partial x^\mu}-y^\theta\frac{\partial^2 G^\mu}{\partial x^\theta\partial y^\mu},
\end{eqnarray}
where
\begin{equation}
2\frac{\partial G^\mu}{\partial x^\mu}=\frac{1}{2}\eta^{\mu\nu}\left(2\frac{\partial^2 h_{\alpha\nu}}{\partial x^\lambda\partial x^\mu}y^\alpha y^\lambda-\frac{\partial^2 h_{\alpha\beta}}{\partial x^\mu\partial x^\nu}y^\alpha y^\beta\right)
\end{equation}
and
\begin{eqnarray}
-y^\theta\frac{\partial^2 G^\mu}{\partial x^\theta\partial y^\mu}=-\frac{y^\theta}{4}\eta^{\mu\nu}\frac{\partial}{\partial x^\theta}\left(2\frac{\partial h_{\mu\nu}}{\partial x^\lambda}y^\lambda+2\frac{\partial h_{\alpha\nu}}{\partial x^\mu}y^\alpha-2\frac{\partial h_{\alpha\mu}}{\partial x^\nu}y^\alpha\right)
-\frac{y^\theta}{4}\frac{\partial \eta^{\mu\nu}}{\partial y^\mu}\frac{\partial}{\partial x^\theta}\bigg(2\frac{\partial h_{\alpha\nu}}{\partial x^\lambda}y^\alpha y^\lambda-\frac{\partial h_{\alpha\beta}}{\partial x^\nu}y^\alpha y^\beta\bigg).
\end{eqnarray}
Since the value of Ricci scaler $R$ is invariant under the coordinate transformation, we must fix the gauge symmetry to yield unique solution. Under a coordinate transformation
\begin{equation}\label{coordinate tranformation}
\bar{x}^\mu=x^\mu+\epsilon^\mu(x),
\end{equation}
the metric $h_{\mu\nu}$ transforms as
\begin{equation}
\bar{h}^{\mu\nu}=h^{\mu\nu}-\frac{\partial \epsilon^\mu}{\partial x^\lambda}\eta^{\lambda\nu}-\frac{\partial \epsilon^\nu}{\partial x^\lambda}\eta^{\lambda\mu}.
\end{equation}
By performing the coordinate transformation with
\begin{equation}
\eta^{\mu\lambda}\frac{\partial^2\epsilon_\nu}{\partial x^\mu \partial x^\lambda}=\frac{\partial h^\mu_{~\nu}}{\partial x^\mu}-\frac{1}{2}\frac{\partial h^{\mu}_{~\mu}}{\partial x^\nu},
\end{equation}
we find that $\bar{h}_{\mu\nu}$ satisfies
\begin{equation}\label{gauge}
\frac{\partial \bar{h}^\mu_{~\nu}}{\partial x^\mu}=\frac{1}{2}\frac{\partial \bar{h}^{\mu}_{~\mu}}{\partial x^\nu}.
\end{equation}
This choice of gauge (\ref{gauge}) has the same form with the Lorentz gauge in general relativity, due to the fact that the locally Minkowshi metric $\eta_{\mu\nu}$ does not depend on $x$.

By making use of the Finslerian gauge (\ref{gauge}), and noticing that $\eta_{\mu\nu}$ does not depend on $x$, we rewrite the Ricci scaler as
\begin{eqnarray}\label{Ricci symplify}
R=-\frac{\eta^{\mu\nu}}{2}\frac{\partial^2 h_{\alpha\beta}}{\partial x^\mu \partial x^\nu}y^\alpha y^\beta
+\frac{1}{4}\frac{\partial \eta^{\mu\nu}}{\partial y^\mu}\frac{\partial^2 h_{\alpha\beta}}{\partial x^\nu \partial x^\lambda}y^\lambda y^\alpha y^\beta.
\end{eqnarray}
We find from (\ref{Ricci symplify}) that the solution of $R=0$ has following properties
\begin{eqnarray}\label{h munu}
h_{\mu\nu}(x,y)=e_{\mu\nu}\exp(ik_\lambda x^\lambda)+h.c.~~~,
\end{eqnarray}
where
\begin{equation}\label{constrain k}
k_\mu k_\nu \eta^{\mu\nu}-\frac{1}{2}\frac{\partial \eta^{\mu\nu}}{\partial y^\mu}k_\nu k_\lambda y^\lambda=0,
\end{equation}
$k=k(y)$ is function of $y$ and $e_{\mu\nu}$ is the polarization tensor.
The term $\frac{\partial \eta^{\mu\nu}}{\partial y^\mu}$ could be written as
\begin{equation}\label{det A}
\frac{\partial \eta^{\mu\nu}}{\partial y^\mu}=-2A^{~\mu\nu}_\mu/\tilde{F}=-\eta^{\nu\lambda}\frac{\partial\ln{\rm |det}(\eta)|}{\partial y^\lambda},
\end{equation}
where $\tilde{F}^2=\eta_{\mu\nu}y^\mu y^\nu$. Substituting the equation (\ref{det A}) into (\ref{constrain k}), we obtain
\begin{equation}\label{dispersion relation}
k_\mu k_\nu \eta^{\mu\nu}=-\eta^{\nu\lambda}\frac{\partial\ln\sqrt{{\rm |det}(\eta)|}}{\partial y^\lambda}k_\nu k_\mu y^\mu~.
\end{equation}
It is obvious that $k_\mu k_\nu \eta^{\mu\nu}\neq0$ while the Finsler spacetime $\eta_{\mu\nu}$ is not Minkowskian. It implies that the wave vectors $k_\mu$ of gravitational waves is not null in Finsler spacetime $\eta_{\mu\nu}$.

The Randers spacetime \cite{Randers} is a special kind of Finsler geometry with Finsler
structure
\begin{equation}
\tilde{F}(x,y)\equiv \alpha+\beta,
\end{equation} where
\begin{eqnarray}
\alpha &\equiv&\sqrt{\bar{a}_{\mu\nu}y^\mu y^\nu},\\
\beta&\equiv& \bar{b}_\mu y^\mu,
\end{eqnarray} and $\bar{a}_{\mu\nu}$ is Riemannian metric. The indices on certain objects that decorated with a bar are lowered and raised by $\bar{a}_{\mu\nu}$ and its matrix inverse $\bar{a}^{\mu\nu}$. Substituting the Randers-Finsler structure $\tilde{F}$ into the dispersion relation of gravitational wave (\ref{dispersion relation}) and supposing the Randers spacetime is very close to Minkowski spacetime $\bar{a}_{\mu\nu}$, to first order in $\bar{b}$, we obtain
\begin{eqnarray}
k_\mu k_\nu \eta^{\mu\nu}&=&-\frac{5(k\cdot\bar{l})}{2}\left((k\cdot\bar{b})-\frac{\beta}{\alpha}(k\cdot\bar{l})\right),\\
k\cdot k&=&-\frac{(k\cdot\bar{l})}{2}\left((k\cdot\bar{b})-\frac{3\beta}{\alpha}(k\cdot\bar{l})\right),
\end{eqnarray}
where $`\cdot'$ denotes the inner product on Minkowski spacetime $\bar{a}_{\mu\nu}$ and $\bar{l}^\mu\equiv y^\mu/\alpha$. The causality should holds in Finsler spacetime $\eta_{\mu\nu}$, thus $k_\mu k_\nu \eta^{\mu\nu}>0$ while the signature of Minkowski metric $\bar{a}_{\mu\nu}$ is of the form $(+~-~-~-)$. If $k\cdot k<0$, it means that the speed of gravitational wave is larger than speed of light. It implies that the speed of gravitational wave could larger than speed of light and causality still holds. The inequalities $k_\mu k_\nu \eta^{\mu\nu}>0$ and $k\cdot k<0$ satisfy if
\begin{equation}\label{requirement}
\frac{3\beta}{\alpha}<\frac{k\cdot\bar{b}}{k\cdot\bar{l}}<\frac{\beta}{\alpha}<0,
\end{equation}
so that the speed of gravitational wave in the anisotropic spacetime is larger than the speed of light and the causality still holds.

The sketch figure of the causal structure of Finsler spacetime ($\eta_{\mu\nu}$) is shown in Fig.1. It is clear from Fig.1 that the null vectors on Finsler spacetime ($\eta_{\mu\nu}$) are spacelike vectors on Minkowski spacetime. The causal speed of Finsler spacetime could be larger than the one of Minkowski spacetime.

\begin{figure}
\includegraphics[scale=1.0]{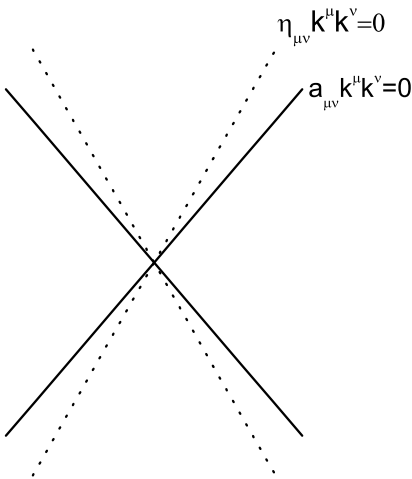}
\caption{The solid line denotes the null structure on Finsler spacetime ($\eta_{\mu\nu}$). The dot line denotes the null structure on Minkowski spacetime. It is obvious that the null structure of Finsler spacetime is larger than the one in Minkowski spacetime}
\end{figure}

\section{Conclusions}
In this paper, we used the weak gravitational field approximation to get a linearized solution of the gravitational vacuum field equation in Finsler spacetime. The plane-waves solution (\ref{h munu}) of gravitational wave in an anisotropic spacetime was presented. It is shown that the gravitational wave is propagating in locally Minkowski spacetime ($\eta_{\mu\nu}$).
The Killing vectors of locally Minkowski spacetime ($\eta_{\mu\nu}$) are investigated in Ref.\cite{Finsler PF}. It was shown that Finsler spacetime admits less symmetry than Minkowski spacetime, and the translation symmetry in preserved in locally Minkowski spacetime ($\eta_{\mu\nu}$). Based on the Noether theorem, the spacetime translational invariance implies that the energy-momentum is well defined and conserved in locally Minkowski spacetime ($\eta_{\mu\nu}$). The dispersion relation of gravitational wave in Finsler spacetime (\ref{dispersion relation}) was presented. The speed of gravitational wave could larger than the speed of light in Randers spacetime and the casuality of gravitational wave still holds, if the condition (\ref{requirement}) is satisfied. Since the wave vector $k_\mu$ of gravitational wave is timelike in locally Minkowski spacetime ($\eta_{\mu\nu}$), it would not lose energy via the gravitational Cherenkov radiation.

\vspace{1cm}
\begin{acknowledgments}
We would like to thank M. H. Li and S. Wang for useful discussions. One of us X. Li thanks Prof. C. Pfeifer for usefol discussions. The work was supported by the NSF of China under Grant No. 10875129 and 11075166 and 11147176.
\end{acknowledgments}

\end{document}